\documentclass[a4paper,12pt]{article}
\usepackage{amsfonts}
\usepackage{latexsym}
\usepackage{amssymb}
\date{}

\begin{document}
%\tableofcontents

\title{Contact Temperature as an Internal Variable\\
of Discrete Systems in Non-Equilibrium\thanks{In memory of G\'{e}rard A. Maugin}
}
\author{W. Muschik\footnote{Corresponding author:
muschik@physik.tu-berlin.de}
\\
Institut f\"ur Theoretische Physik\\
Technische Universit\"at Berlin\\
Hardenbergstr. 36\\D-10623 BERLIN,  Germany}
\maketitle

            \newcommand{\be}{\begin{equation}}
            \newcommand{\beg}[1]{\begin{equation}\label{#1}}
            \newcommand{\ee}{\end{equation}\normalsize}
            \newcommand{\bee}[1]{\begin{equation}\label{#1}}
            \newcommand{\bey}{\begin{eqnarray}}
            \newcommand{\byy}[1]{\begin{eqnarray}\label{#1}}
            \newcommand{\eey}{\end{eqnarray}\normalsize}
            \newcommand{\beo}{\begin{eqnarray}\normalsize}
            \newcommand{\R}[1]{(\ref{#1})}
            \newcommand{\C}[1]{\cite{#1}}

            \newcommand{\mvec}[1]{\mbox{\boldmath{$#1$}}}
            \newcommand{\x}{(\!\mvec{x}, t)}
            \newcommand{\m}{\mvec{m}}
            \newcommand{\F}{{\cal F}}
            \newcommand{\n}{\mvec{n}}
            \newcommand{\argm}{(\m ,\mvec{x}, t)}
            \newcommand{\argn}{(\n ,\mvec{x}, t)}
            \newcommand{\T}[1]{\widetilde{#1}}
            \newcommand{\U}[1]{\underline{#1}}
            \newcommand{\V}[1]{\overline{#1}}
            \newcommand{\X}{\!\mvec{X} (\cdot)}
            \newcommand{\cd}{(\cdot)}
            \newcommand{\Q}{\mbox{\bf Q}}
            \newcommand{\p}{\partial_t}
            \newcommand{\z}{\!\mvec{z}}
            \newcommand{\bu}{\!\mvec{u}}
            \newcommand{\rr}{\!\mvec{r}}
            \newcommand{\w}{\!\mvec{w}}
            \newcommand{\g}{\!\mvec{g}}
            \newcommand{\D}{I\!\!D}
            \newcommand{\se}[1]{_{\mvec{;}#1}}
            \newcommand{\sek}[1]{_{\mvec{;}#1]}}            
            \newcommand{\seb}[1]{_{\mvec{;}#1)}}            
            \newcommand{\ko}[1]{_{\mvec{,}#1}}
            \newcommand{\ab}[1]{_{\mvec{|}#1}}
            \newcommand{\abb}[1]{_{\mvec{||}#1}}
            \newcommand{\td}{{^{\bullet}}}
            \newcommand{\eq}{{_{eq}}}
            \newcommand{\eqo}{{^{eq}}}
            \newcommand{\f}{\varphi}
            \newcommand{\rh}{\varrho}
            \newcommand{\dm}{\diamond\!}
            \newcommand{\seq}{\stackrel{_\bullet}{=}}
            \newcommand{\st}[2]{\stackrel{_#1}{#2}}
            \newcommand{\om}{\Omega}
            \newcommand{\emp}{\emptyset}
            \newcommand{\bt}{\bowtie}
            \newcommand{\btu}{\boxdot}
            \newcommand{\tup}{_\triangle}
            \newcommand{\tdo}{_\triangledown} 
\newcommand{\Section}[1]{\section{\mbox{}\hspace{-.6cm}.\hspace{.4cm}#1}}
\newcommand{\Subsection}[1]{\subsection{\mbox{}\hspace{-.6cm}.\hspace{.4cm}
\em #1}}

\newcommand{\const}{\textit{const.}}
\newcommand{\vect}[1]{\underline{\ensuremath{#1}}}  %Vektoren
\newcommand{\abl}[2]{\ensuremath{\frac{\partial #1}{\partial #2}}}

\noindent
{\bf Abstract}\ State space and entropy rate of a discrete non-equilibrium system are shortly
considered including internal variables and the contact temperature. The concept of internal
variables in the context of non-equilibrium thermodynamics of a closed discrete system is 
discussed. The difference between internal variables and degrees of freedom are repeated, and
different types of their evolution equations are mentioned in connection with G\'{e}rard A.
Maugin's numerous papers on applications of internal variables. The non-equilibrium contact
temperature is recognized as an internal variable and its evolution equation is presented.

\section{Introduction}

Temperature is a quantity which can be measured easily, but whose theoretical background is
complicated. There is a huge variety of different thermometers \C{23,1} all measuring "temperature",
but
the concept of temperature is first of all only properly defined in equilibrium. For elucidating this fact,
we consider the simple example of a thermometer whose surface\footnote[1]{The thermometer is here
a discrete system which has a volume and a surface, as small as ever.} has different heat
conductivities. Contacted with a non-equilibrium system, the measured "temperature" depends
at the same position on the orientation of such a "thermometer". Clear is, that this orientation
sensitivity of the thermometer vanishes in equilibrium (if one knows what equilibrium is). A second
example is a thermometer which measures the intensity of radiation which is composed of different
parts of the spectrum. The measured "temperature" depends on the sensitivity distribution of the
thermometer over the spectrum with the result, that different thermometers measure different
"temperatures" at the same object.
%\vspace{.3cm}\newline%%%%%%%%%%%%%%%%%%%%%%%%%%%%%%%

For escaping these "thermometer induced" difficulties, a theoretical definition of temperature is
considered as a remedy . We define \C{1}
\bee{B1}
\mbox{discrete systems:}\quad\frac{1}{T}\ :=\ \frac{\partial S}{\partial U},\qquad
\mbox{field formulation:}\quad\frac{1}{T}\ :=\ \frac{\partial s}{\partial u}.
\ee
But also these definitions have their malices: First of all, a state space is needed, because the
partial derivatives have no sense without it. Then entropy $S$ or entropy density $s$  and internal
energy
$U$ or internal energy density $u$ are needed in equilibrium or out of it. And finally, the open question is,
if there exists a thermometer which measures the temperature $T$. 
%\vspace{.3cm}\newline%%%%%%%%%%%%%%%%%%%%%%%%%%%%%%%

To avoid all these uncertainties, a simple idea is the following: why not define a general concept
of temperature which is valid independently of equilibrium or non-equi\-li\-brium and which is
introduced into the theoretical framework by defining the RHSs of \R{B1}
\bee{B2}
\frac{1}{\Theta}\ =:\ \frac{\partial S}{\partial U},\qquad
\frac{1}{\Theta}\ =:\ \frac{\partial s}{\partial u}\ ?
\ee
If additionally $\Theta$ is connected with a measuring instruction which "defines" the temperature
$\Theta$ experimentally, temperature comes from the outside into the theoretical framework and
not vice versa. How to realize this idea in connection with internal variables is the intension of this paper.

\section{Contact temperature}
\subsection{Definition}

We consider a closed discrete non-equilibrium system\footnote[2]{For the
sake of a minimum of formalism, we consider here closed systems. This choice does not influence
the definition of the contact temperature below, that means, closing an open system does not
change its contact temperature. More details in \C{2}.} which is contacted with an equilibrium
environment of thermostatic temperature $T^*$. The heat exchange per time between the
considered system and its environment is $\st{\td}{Q}$.
We now introduce a temperature $\Theta$ which satisfies the inequality
\bee{B3}
\st{\td}{Q}\Big(\frac{1}{\Theta}-\frac{1}{T^*}\Big)\ \geq\ 0.
\ee
According to this inequality, we obtain\footnote[3]{A more detailled proof is represented in
App.\ref{HC}}
\bee{B4}
\st{\td}{Q}\ > 0 \ \Longrightarrow \ T^* > \Theta, \qquad
\st{\td}{Q}\ < 0 \ \Longrightarrow \ T^* < \Theta,\qquad
T^* = \Theta\ \ \Longleftrightarrow\ \ \st{\td}{Q}\, = 0.
\ee
Consequently, we have the following
\vspace{.3cm}\newline
{\sf Definition:}
\begin{center}
\parbox{12cm}{The system's contact temperature is that thermostatic temperature
of the system's equilibrium environment for which the net heat echange between
the system and this environment through an inert\footnotemark[4] partition
vanishes by change of sign\footnotemark[5].}
\end{center}
\footnotetext[4]{inert means: the partition does not absorb or emit energy
and/or material}
\footnotetext[5]{Do not take the vanishing net heat
exchange for an adiabatic condition: there are positive and negative heat
exchanges through partial surfaces between system and environment.}
The contact temperature is defined for discrete systems in non-equilibrium embracing the case of
equilibrium \C{3,4,5}. In  both cases, the net heat exchange vanishes, if the thermostatic temperature of the
controlling environment $T^*$ is equal to the contact temperature $\Theta$ in non-equilibrium
according to \R{B4}$_3$, or if $T^*$ is equal to the thermostatic temperature $T$ of the system
in equilibrium. If the system is in non-equilibrium at the contact temperature $\Theta = T^*$,
the sum of the non-vanishing partial heat exchanges between system and heat reservoir of $T^*$
vanishes. If the system is in equilibrium, all these partial heat exchanges vanish.  
%\vspace{.3cm}\newline%%%%%%%%%%%%%%%%%%%%%%%%%%%%%%

The contact temperature is not defined by \R{B2}, but it is a basic quantity similar as the energy.
Entropy and internal energy in connection with a suitable state space have to be defined so that
\R{B2} is satisfied. This item is treated in sect.\ref{SS}.

\subsection{Contact temperature and internal energy\label{CT}}

The contact temperature $\Theta$ is independent of the internal energy $U$ of the
system\footnote[6]{see \C{6} 4.1.2}. 
The proof of this statement runs as follows: we consider the energy balance equation of a closed
discrete system with a rigid power impervious partition
\bee{B5}
\st{\td}{U}\ =\ \st{\td}{Q} +\st{\td}{W},\qquad\st{\td}{W}\ =\ 0,\ \mbox{the power}.
\ee
The process taking place in the non-equilibrium system generates a time dependent contact temperature
$\Theta(t)$ and a time dependent heat exchange $\st{\td}{Q}\!(t)$ which also depends on the
tem\-pe\-ra\-ture $T^*(t)$ of the equilibrium environment which controls the system. 
\vspace{.3cm}\newline
We now choose the environment's temperature for all times equal to the contact temperature of the
system, and we obtain
\bee{B6}
T^*(t)\ \doteq\ \Theta(t)\quad \Longrightarrow\quad \st{\td}{Q}\!(t)\ =\ 0\quad
\Longrightarrow\quad \st{\td}{U}(t)\ =\ 0.
\ee
The last implication is due to \R{B5}. Because $\Theta$ is time dependent and $U$ is constant,
both quantities are independent of each other.

\section{State Space and Entropy Rate\label{SS}}

We consider the state space of a closed discrete non-equilibrium system which contains the contact
temperature as an independent variable according to sect.\ref{CT}
\bee{B7}
Z\ :=\ (U,\mvec{a},\Theta,\mvec{\xi}).
\ee
Here, the $\mvec{a}$ are the work variables
\bee{B8}
\st{\td}{W}\ =\ {\bf A}\cdot\st{\td}{\mvec{a}}
\ee
and $\mvec{\xi}$ the internal variables --"measurable but not controllable"\footnote[7]{see \C{7} 4.1;
\C{8} 5.6}-- which are discussed in sect.\ref{IV}. 
%\vspace{.3cm}\newline%%%%%%%%%%%%%%%%%%%%%%%%%%%%%%

A process $Z(t)$ is represented by a trajectory $\cal T$ on the non-equilibrium space \R{B7}.
According to this state space, the time rate of the non-equilibrium entropy becomes
along $\cal T$ by inserting the first law \R{B5}
\bee{B9}
\st{\td}{S}\ :=\ \frac{1}{\Theta}\Big(\st{\td}{U}-{\bf A}\cdot\st{\td}{\mvec{a}}\Big)
+\alpha\st{\td}{\Theta}+\mvec{\beta}\cdot\st{\td}{\mvec{\xi}}\ =\
\frac{1}{\Theta}\st{\td}{Q}+\alpha\st{\td}{\Theta}+\mvec{\beta}\cdot\st{\td}{\mvec{\xi}}.
\ee
By definition, the entropy rate of an isolated system is the entropy
production $\Sigma$ which is non-negative according to the second law
\bee{B10}
\st{\td}{U}\ \equiv\ 0,\quad \st{\td}{\mvec{a}}\ \equiv\ 0\ \longrightarrow\
\st{\td}{S}\!{^{isol}}\ =:\ 
\Sigma\ =\ \alpha\st{\td}{\Theta}+\mvec{\beta}\cdot\st{\td}{\mvec{\xi}}\ \geq\ 0.
\ee
Because the contact temperature is independent of the other internal variables, we can decompose
\R{B10}$_4$
\bee{B10a}
\alpha \st{\td}{\Theta}\ \geq\ 0,\qquad\mvec{\beta}\cdot\st{\td}{\mvec{\xi}}\ \geq\ 0,
\ee
and using \R{B9} and \R{B3}, we obtain the inequalities
\bee{B10b}
\st{\td}{S}\ \geq\ \frac{\st{\td}{Q}}{\Theta}\ \geq\ \frac{\st{\td}{Q}}{T^*}.
\ee
%\vspace{.3cm}\ee%%%%%%%%%%%%%%%%%%%%%%%%%%%%%%%%%%%
%\vspace{.3cm}\newline%%%%%%%%%%%%%%%%%%%%%%%%%%%%

If we presuppose that a state function $S(U,\mvec{a},\Theta,\mvec{\xi})$
exists\footnote[8]{That is
the case in large state spaces, if the system is adiabatically unique \C{2}.}, the integration along a
cyclic trajectory on \R{B7} results in the extended Clausius inequality of closed systems
\bee{B10c}
0\ \geq\ \oint\frac{\st{\td}{Q}}{\Theta}dt\ \geq\ \oint\frac{\st{\td}{Q}}{T^*}dt,
\ee
and \R{B2}$_1$ becomes an integrability condition
\bee{B11}
\frac{1}{\Theta}\ =\ \Big(\frac{\partial S}{\partial U}\Big)_{\mvec{a},\Theta,\mvec{\xi}}\quad
\longrightarrow\quad
S(U,\mvec{a},\Theta,\mvec{\xi})\ =\ \frac{1}{\Theta}U+K(\mvec{a},\Theta,\mvec{\xi}),
\ee
and $-\Theta K(\mvec{a},\Theta,\mvec{\xi})$ is the free energy\footnote[9]{more details in \C{2,18}}.
Because of the two last terms in \R{B9}, the contact temperature takes the placing of an internal
variable which are discussed in sect.\ref{IV}.

\section{Equilibrium and Reversible "Processes"}

Equilibrium in thermally homogeneous systems satisfies the following equilibrium conditions:
\byy{B12}
\mbox{no time dependence:}& & \st{\td}{\boxtimes}{^{eq}}\ \doteq\ 0,\hspace{3cm}
\\ \label{B13}
\mbox{thermostatic temperature:}& & \Theta^{eq}\ \doteq\ T(U,\mvec{a})\ =\ T^*.
\eey
Consequently, we obtain
\bee{B14}
\st{\td}{U}{^{eq}}\ =\ 0,\quad \st{\td}{\mvec{a}}{^{eq}}\ =\ \mvec{0},\quad 
\st{\td}{Q}{^{eq}}\ =\ 0,\quad \st{\td}{\Theta}{^{eq}}\ =\ 0,\quad
\st{\td}{\mvec{\xi}}{^{eq}}\ =\ \mvec{0}\longrightarrow\ \Sigma^{eq}\ =\ 0
\ee
according to \R{B10}.
We now have to distinguish two kinds of equilibria concerning the affinities $\mvec{\beta}$
\bee{B15}
\mvec{\beta}\Big(U,\mvec{a},T(U,\mvec{a}),\mvec{\xi^{eq}}\Big)\ =\
\mvec{0}\quad\mbox{and}\quad
\mvec{\beta}\Big(U,\mvec{a},T(U,\mvec{a}),\mvec{\xi}_f\Big)\ \neq\ \mvec{0}.
\ee
Consequently, we obtain
\byy{B16}
\mbox{unconstraint equilibrium:}&\quad&
\st{\td}{\mvec{\xi}}{^{eq}}\ =\ \mvec{0}\ \wedge\ \mvec{\beta}\ =\ \mvec{0},
\\ \label{B17}
\mbox{constraint equilibrium:}&\quad&
\st{\td}{\mvec{\xi}}_f\ =\ \mvec{0}\ \wedge\ \mvec{\beta}\ \neq\ \mvec{0}.
\eey
Unconstraint equilibrium means that the internal variables are fixed at their equilibrium values
\bee{B18}
\mvec{\xi^{eq}}\ =\ \mvec{\xi}(U,\mvec{a})\ \neq\ \mvec{\xi}_f
\ee
according to the solution of \R{B15}$_1$. Constraint equilibrium means that the internal variables
$\mvec{\xi}_f$ are according to \R{B17}$_1$ "frozen in" at a value which is different from its
unconstraint equilibrium value and which is not determined by $(U,\mvec{a})$. "Mixed equilibria" are
possible in which one part $\mvec{\zeta}$ of the internal variables is in unconstraint equilibrium, whereas the other part $\mvec{\chi}$ of them is in constraint equilibrium:
\bee{B18a}
\mvec{\xi}\ =\ (\mvec{\zeta},\mvec{\chi})\ \longrightarrow\ 
\mvec{\xi}^{eq}\ =\ \Big(\mvec{\zeta}^{eq}(U,\mvec{a}),\mvec{\chi}^{eq}=\mvec{\chi}_f\Big).
\ee
The individual parts $\mvec{\zeta}$ and $\mvec{\chi}$ of a mixed equilibrium may depend on the
time for which the system is isolated making the relaxation to equilibrium possible.
%\vspace{.3cm}\newline%%%%%%%%%%%%%%%%%%%%%%%%%%%%

The dimension of the state space \R{B7} shrinks at equilibrium according to \R{B13}
and \R{B18a}$_1$
\bee{B19}
Z^{eq}\ :=\ \Big(U,\mvec{a},\Theta^{eq}(U,\mvec{a}),\mvec{\zeta}^{eq}(U,\mvec{a}),
\mvec{\chi_f}\Big)\ \longrightarrow\ {\cal Z}^{eq}\ =\ (U,\mvec{a},\mvec{\chi}_f).
\ee
%\vspace{.3cm}\ee%%%%%%%%%%%%%%%%%%%%%%%%%%%%%%%%%

Such as the entropy rate of a non-equilibrium process $\cal T$ is defined on \R{B7}, we define the
reversible entropy "rate" along a {\em reversible "process"} $\cal R$ on the equilibrium sub-space
\R{B19}$_2$ ${\cal Z}^{eq}(t)$
\bee{B20}
\st{\td}{S}{^{rev}}\ :=\ \frac{1}{T}\Big(\st{\td}{U}- {\bf A}^{rev}\cdot\st{\td}{\mvec{a}}\Big)
+\frac{\partial S^{rev}}{\partial\mvec{\chi}_f}\cdot \st{\td}{\mvec{\chi}_f},\qquad
\st{\td}{U}- {\bf A}^{rev}\cdot\st{\td}{\mvec{a}}\ =:\ \st{\td}{Q}{^{rev}}.
\ee
The process parameter "$t$" along $\cal R$ is not the real time because real processes are not
possible in
the equilibrium sub-space. The time parameter is formally generated by projection of $\cal T$ onto $\cal R$ 
\bee{B21}
{\cal P}{\cal T}(t)\ =\
{\cal P}(U,\mvec{a}, \Theta,\mvec{\xi})(t)\ =\ (U,\mvec{a},\mvec{\chi}_f)(t)\ =\
{\cal R}(t),
\ee
where the reversible {\em accompanying "process"} $(U,\mvec{a},\mvec{\chi}_f)(t)$ takes
place\footnote[10]{$t$ is the "slaved time" according to \R{B21}}
which belongs to the original one $(U,\mvec{a}, \Theta,\mvec{\xi})(t)$ by projection \C{CISM}.
%\vspace{.3cm}\newline%%%%%%%%%%%%%%%%%%%%%%%%%%%%%%%%%

To connect \R{B9} with \R{B20}$_1$, we apply the {\em embedding axiom} \C{6}
\bee{B22}
S_{B(eq)}-S_{A(eq)}\ =\ {\cal T}\int_{A}^{B}\st{\td}{S}(t)dt\ \doteq\
{\cal R}\int_{A}^{B}\st{\td}{S}{^{rev}}(t)dt
\ee
which by use of \R{B9}$_2$ and \R{B20} results in
\bee{B23}
({\cal T}/{\cal R})\int_{A}^{B}\Big(\frac{\st{\td}{Q}}{\Theta}-\frac{\st{\td}{Q}{^{rev}}}{T}
+ \alpha\st{\td}{\Theta}+\mvec{\beta}\cdot\st{\td}{\mvec{\xi}}
-\frac{\partial S^{rev}}{\partial\mvec{\chi}_f}\cdot \st{\td}{\mvec{\chi}_f}\Big)dt\ =\ 0.
\ee
We obtain according to \R{B3} and \R{B20}$_2$, and that $T(U,\mvec{a})=T^*$
is valid along ${\cal R}$ 
\bee{B23a}
\frac{\st{\td}{Q}}{\Theta}-\frac{\st{\td}{Q}{^{rev}}}{T}\ \geq\ 
\frac{\st{\td}{Q}}{T^*}-\frac{\st{\td}{Q}{^{rev}}}{T}\ =\
\frac{1}{T^*}\Big(\st{\td}{Q}-\st{\td}{Q}{^{rev}}\Big)\ =\
\frac{1}{T^*}\Big({\bf A}^{rev}-{\bf A}\Big)\cdot\st{\td}{\mvec{a}}.
\ee
Taking \R{B10}$_3$ and \R{B23a} into account, we obtain from \R{B23}
\bee{B24} 
{\cal T}/{\cal R}\int_{A}^{B}\Big[
\frac{1}{T^*}\Big({\bf A}^{rev}-{\bf A}\Big)\cdot\st{\td}{\mvec{a}}
-\frac{\partial S^{rev}}{\partial\mvec{\chi}_f}\cdot \st{\td}{\mvec{\chi}_f}\Big]dt\
\leq\ 0.
\ee
If the system under consideration has only unconstraint equilibria, \R{B24} yields
\bee{B24a}
{\cal R}\int_{A}^{B}{\bf A}^{rev}\cdot\st{\td}{\mvec{a}}dt\ \leq\ 
{\cal T}\int_{A}^{B}{\bf A}\cdot\st{\td}{\mvec{a}}dt,
\ee
and we obtain for the volume work the well known inequality
\bee{B24b}
{\bf A}\cdot\st{\td}{\mvec{a}}\ \equiv\ -p\st{\td}{V}\ \longrightarrow\
{\cal R}\int_{A}^{B}p^{rev}dV\ \geq\ 
{\cal T}\int_{A}^{B}p\st{\td}{V}dt.
\ee
%\vspace{.3cm}\ee%%%%%%%%%%%%%%%%%%%%%%%%%%%%%%%%%

The non-equilibrium entropy \R{B9} has to be complemented by evolution laws for the internal
variables $\Theta$ and $\mvec{\xi}$ in the next section.

\section{Brief View at Internal Variables\label{IV} }

Historically, the concept of internal variables can be traced back to Bridgman \C{16}, Meixner
\C{19} and many others. The introduction of internal variables makes possible to use large state
spaces, that means,
material properties can be described by mappings defined on the state space variables
(including the internal ones), thus avoiding the use of their histories which appear in
small state spaces \C{17}. Those are generated, if the internal variables are eliminated.
Consequently, internal variables allow to use the methods of Irreversible and/or Extended
Thermodynamics \C{7}.
%\vspace{.3cm}\newline%%%%%%%%%%%%%%%%%%%%%%%%%%%%%%%%

Internal variables cannot be chosen arbitrarily: there are seven concepts which restrict their
introduction \C{17}. The most essential ones are:\newline
(i) Internal variables need a model or an interpretation,\newline
(ii) Beyond the constitutive and balance equations, internal variables require rate
equations which can be adapted to different situations, making the use of internal variables flexible
and versatile,\newline
(iii) The time rates of the internal variables do not occur in the work differential of the First Law,
\newline
(iv) An isolation of the discrete system does not influence the internal variables,\newline
(v) In equilibrium, the internal variables become dependent on the variables of the equilibrium sub-space, if the equilibrium is unconstraint.
%\vspace{.3cm}\newline%%%%%%%%%%%%%%%%%%%%%%%%%%%%%%%%

Satisfying these concepts, the internal variables entertain an ambiguous relationship with
microstructure and internal degrees of freedom \C{15}. But internal variables and internal degrees
of freedom represent different concepts for extending the state space: both are included in the
state space, both need evolution laws, but whereas internal variables do not occur in the work
differential of the First Law according to (iii), degrees of freedom appear in the time rate of the
internal energy. Consequently, the question "internal variables or degrees of freedom ?"\footnote[11]{This question was discussed during G\'{e}rard's stay at the Wissenschaftskolleg zu Berlin, 1991/92, \C{7} sect.4.7 B, \C{15}} can be answered clearly. %\vspace{.3cm}\newline%%%%%%%%%%%%%%%%%%%%%%%%%%%%%%%%

As the last term of \R{B9} shows, internal variables must be complemented by an evolution
law\footnote[12]{\C{7}, 3.5, 4.7.B, \C{15}I} which may have the shape
\bee{B25}
\st{\td}{\mvec{\xi}}\ =\ \mvec{f}(U,\mvec{a},\mvec{\xi})
+ \mvec{g}(U,\mvec{a},\mvec{\xi})\st{\td}{U}
+\  {\bf h}(U,\mvec{a},\mvec{\xi})\cdot\st{\td}{\mvec{a}}.
\ee
Special one-dimensional cases are
\byy{B26}
\mbox{relaxation type:}&&\quad \st{\td}{\xi}(t)\ =\ -\frac{1}{\tau(U,\mvec{a},\Theta)}\Big(\xi(t)
-\xi^{eq}\Big),
\\ \label{B27}
\mbox{reaction type \C{17}:}&&\quad  \st{\td}{\xi}(t)\ =\ \gamma(U,\mvec{a},\Theta)\Big[
1-\exp\Big(-\mu(t)\beta(U,\mvec{a},\Theta)\Big)\Big].
\eey
%\vspace{.3cm}\eey%%%%%%%%%%%%%%%%%%%%%%%%%%%%%%%%%%%%

If for a special degree of freedom an evolution criterion exists \C{22}
\bee{B28}
\frac{d}{dt}\int_{G(t)}{\cal L}(...)dV\ \geq\ 0,
\ee
we obtain a variational problem at equilibrium
\bee{B29}
\Big(\int_{G(t)}{\cal L}(...)dV\Big)^{eq}\ \longrightarrow\ \mbox{max},
\ee
and the Euler-Lagrange equations of $\cal L$ result in the Landau-Ginzburg equations for the
considered degree of freedom at equilibrium\footnote[13]{An example for a degree of freedom is the
second order alignment tensor of liquid crystal theory \C{22}.}.

\section{Contact Temperature as an Internal Variable }

When at the University of Calgary (Canada) on the first week of August, 1979, G\'{e}rard delivered a lecture on
"Electromagnetic internal variables in ferroelectric and ferromagnetic continua", he started a series
of papers on internal variables \C{9} -\C{20} which comes to its end in 2013 \C{21}. 
The concept of contact temperature is mentioned in sect.4.3 of \C{7}, but without any connection to
internal variables. In the sequel, G\'{e}rard replaced the contact temperature in the Clausius-Duhem
inequality by the thermostatic temperature, thus blurring the differences between these two concepts
of temperature. Beyond that, nobody was aware at that times, that despite of its appearance in the entropy rate \R{B9} the contact temperature may be an internal variable . This knowledge came into
consideration in the course of 2012/14 \C{18}. 
%\vspace{.3cm}\newline%%%%%%%%%%%%%%%%%%%%%%%%%%%%%%%%

Starting out with the defining inequality of the contact temperature \R{B3}, we obtain the following
constitutive equation
\bee{B30}
\st{\td}{Q}\ =\ \kappa(T^*-\Theta),\quad\kappa\ =\ {\cal K}[T^*-\Theta],\quad\kappa\ >\ 0.
\ee
The heat exchange number $\kappa$ is positive and depends on the temperature difference between
the non-equilibrium system of contact temperature $\Theta$ and the equilibrium environment of
thermostatic temperature $T^*$.
%\vspace{.3cm}\newline%%%%%%%%%%%%%%%%%%%%%%%%%%%%%%%%

We now explain in four steps why the contact temperature is an internal variable \C{18}:\newline
{\bf 1:} We consider the pure thermal contact between the non-equilibrium system and its
equilibrium environment --marked by $^*$-- through an inert partition. The First Law of the equilibrium environment --the heat reservoir-- is
\bee{B31}
\st{\td}{U}{^*} =\ \st{\td}{Q}{^*} =\ -\st{\td}{Q},
\qquad\st{\td}{\mvec{a}}{^*}\ \equiv\ \mvec{0}.
\ee
The caloric equation of the heat reservoir is
\bee{B32}
T(U^*,\mvec{a}^*)\ =\ T^*\ \longrightarrow\
\frac{\partial T}{\partial U^*}(-\st{\td}{Q})\ =\ \st{\td}{T}{^*}\ \longrightarrow\
\st{\td}{Q}\ =\ -\frac{\st{\td}{T}{^*}}{\partial T/\partial U^*}.
\ee
Consequently, the heat exchange can be measured by calorimetry using the heat reservoir.
\newline
{\bf 2:} If according to \R{B4}$_3$, the net heat exchange between the non-equilibrium
system and the heat reservoir vanishes, the non-equilibrium systems has by definition
the contact temperature $\Theta$. Consequently, the contact temperature is measurable, but not
contollable by the heat reservoir.\newline
{\bf 3:} Because $T^*$, $\st{\td}{Q}$ and $\Theta$ are measurable quantities , also the
function $\cal K$ is according to \R{B30}$_1$ known by measurement.\newline
{\bf 4:} We obtain from \R{B30} the time rate of the heat exchange
\bee{B33} 
\p\! \st{\td}{Q}\ =\ \Big({\cal K}'[T^*-\Theta](T^*-\Theta)+\kappa\Big)
(\st{\td}{T}{^*}-\st{\td}{\Theta})
\ee
resulting in an evolution equation for the contact temperature
\bee{B34}
\st{\td}{\Theta}\ =\ \st{\td}{T}{^*} 
- \frac{\kappa\p\! \st{\td}{Q}}{{\cal K}'\st{\td}{Q}+\kappa^2}.
\ee
Thus, because of \#{\bf 2} and \R{B34}, the contact temperature $\Theta$ is an internal variable. 
%\vspace{.3cm}\newline%%%%%%%%%%%%%%%%%%%%%%%%%%%%%%%%

For the sake of simplicity, the special case of a closed discrete system is here considered. The
general case of open discrete systems and of field formulation is treated in \C{18}, a paper
which was dedicated to G\'{e}rard on the occasion of his 70th birthday in 2014\footnote[14]{Other
papers dedicated to him are \C{24} and \C{25}.}.

\section{Appendices}
\subsection{Heat exchange and contact temperature\label{HC}}

The heat exchange $\st{\td}{Q}$ between the considered non-equilibrium system and the equilibrium environment of the
thermostatic temperature $T^*$ represent a two-place one-to-one correlation $\cal R$ satisfying
the two statements
\byy{A1} 
(T^*,\st{\td}{Q})\in{\cal R}\ \wedge\ (T^*,\st{\td}{Q}_0)\in{\cal R} &\Longrightarrow&
\st{\td}{Q}\ =\ \st{\td}{Q}_0,
\\ \label{A2}
(T^*,\st{\td}{Q})\in{\cal R}\ \wedge\ (T^*_0,\st{\td}{Q})\in{\cal R} &\Longrightarrow&
T^*\ =\ T^*_0.
\eey
In a more physical diction this means: To each temperature of the environment belongs a unique
heat exchange between system and environment and vice versa.
%\vspace{.3cm}\newline%%%%%%%%%%%%%%%%%%%%%%%%%%%%%%%%%%

We now introduce a temperature $\Theta$ with the property
\byy{A3} 
(T^*,\st{\td}{Q})\in{\cal R}\ \wedge\ \st{\td}{Q}\ \geq\ 0 &\Longrightarrow& \Theta\leq T^*,
\\ \label{A4}
(T^*,\st{\td}{Q})\in{\cal R}\ \wedge\ \Theta\geq T^*  &\Longrightarrow& \st{\td}{Q}\ \leq\ 0.
\eey
Especially for $T^*=\Theta$ follows
\byy{A5} 
(\Theta,\st{\td}{Q})\in{\cal R}\ \wedge\ \st{\td}{Q}\ \geq\ 0 &\Longrightarrow& \Theta =\Theta,
\\ \label{A6}
(\Theta,\st{\td}{Q})\in{\cal R}\ \wedge\ \Theta = \Theta &\Longrightarrow& \st{\td}{Q}\ \leq\ 0,
\eey
resulting in
\bee{A6}
(\Theta,0)\in{\cal R}
\ee
which means: the heat exchange $\st{\td}{Q}$ vanishes, if the environment has the {\em contact
temperature} $\Theta$ and vice versa because of the one-to-one correlation $\cal R$. Although the contact temperature is by definition a thermostatic one of the equilibrium environment, we
attach it to the non-equilibrium system as a non-equilibrium temperature which satisfy the defining inequality \R{B3}. Of course, the value of the contact temperature depends on the properties of
the partition generating the contact between non-equilibrium system and heat reservoir. The
denotation "contact temperature" stems from this contact depending which disappears in
equilibrium.

\subsection{Contact temperature and efficiency}

We consider a cyclic, power-producing process of a closed discrete non-equilibrium system
which works between two heat reservoirs of constant thermostatic temperatures
$T^*_H  > T^*_L$ \C{2}.
The contact temperatures of the two contacts between the
system and the corresponding reservoirs are $\Theta_H(t)$ and $\Theta_L(t)$,
the heat exchanges through the inertial contacts are
$\st{\td}{Q}\!{^*_H}(t)<0 $ and $\st{\td}{Q}\!{^*_L}(t)>0$, relative to the heat reservoirs.
According to the defining inequality \R{B3}, we obtain for the two heat reservoirs
\bee{A7}
\Big(\frac{1}{T^*_H }-\frac{1}{\Theta_H}\Big)\st{\td}{Q}{^*_H}\ \geq\ 0,\qquad
\Big(\frac{1}{T^*_L }-\frac{1}{\Theta_L}\Big)\st{\td}{Q}{^*_L}\ \geq\ 0,
\ee
resulting in
\bee{A8}
T^*_H\ \geq\  \Theta_H,\qquad\Theta_L\ \geq\ T^*_L .
\ee
Integration over the cycle time yields
\byy{A9}
\oint\frac{\st{\td}{Q}{^*_H}}{\Theta_H}dt&\leq& \frac{1}{T^*_H}\oint\st{\td}{Q}{^*_H}dt\
=:\ \frac{1}{T^*_H}Q^*_H,
\\ \label{A10}
\oint\frac{\st{\td}{Q}{^*_L}}{\Theta_L}dt&\leq& \frac{1}{T^*_L}\oint\st{\td}{Q}{^*_L}dt\
=:\ \frac{1}{T^*_L}Q^*_L.
\eey
%\vspace{.3cm}\eey%%%%%%%%%%%%%%%%%%%%%%%%%%%%%%%%%%%

The mean value theorem applied to \R{A9}$_1$ and \R{A10}$_1$ results in
\byy{A11}
\oint\frac{\st{\td}{Q}{^*_H}}{\Theta_H}dt&=&\frac{Q^*_H}{[\Theta_H]}\ \leq\
\frac{1}{T^*_H}Q^*_H\ \longrightarrow\ T^*_H\ \geq\ [\Theta_H],
\\ \label{A12}
\oint\frac{\st{\td}{Q}{^*_L}}{\Theta_L}dt&=&\frac{Q^*_L}{[\Theta_L]}\ \leq\
\frac{1}{T^*_L}Q^*_L\ \longrightarrow\ T^*_L\ \leq\ [\Theta_L],
\eey
Here, the square brackets denote mean values over the cyclic process which are defined by
\R{A11}$_1$ and \R{A12}$_1$. We obtain the following
estimation of the Carnot efficiency according to \R{A11}$_3$ and \R{A12}$_3$
\bee{A13}
\eta_{Car}\ =\ 1-\frac{T^*_L}{T^*_H}\ \geq\ 1-\frac{[\Theta_L]}{[\Theta_H]}\ =:\ \eta_{neq}.
\ee
The non-equilibrium efficiency $\eta_{neq}$ is smaller or equal to the Carnot efficiency which
belongs to reversible processes in contrast to $\eta_{neq}$ which is a more realistic efficiency.
\vspace{.3cm}\newline
An essential presupposition for the considerations above is according to \R{A8}
\bee{A14}
T^*_H\ \geq\  \Theta_H(t),\qquad\Theta_L(t)\ \geq\ T^*_L, 
\ee
that the contact temperatures during the non-equilibrium process satisfy \R{A14} for all times.
%\vspace{.3cm}\newline%%%%%%%%%%%%%%%%%%%%%%%%%%%%%%%%

We now consider the First and the Second Law with respect to the contact temperature. The First Law
per cycle of the considered power-producing non-equilibrium system runs as follows
\bee{A15}
Q_H + Q_L + W = 0\quad\longrightarrow\quad Q^*_H + Q^*_L = W < 0
\quad\longrightarrow\quad Q^*_H = W- Q^*_L < - Q^*_L,
\ee
and the Clausius inequality (Second Law) \R{B10c}$_1$ becomes by use of the mean value theorem
\bee{A16}
 0\ \leq\oint\frac{\st{\td}{Q^*}}{\Theta}dt\ =\ \oint\frac{\st{\td}{Q}\!{^*_H}}{\Theta_H}dt
+\oint\frac{\st{\td}{Q}\!{^*_L}}{\Theta_L}dt\ =\ \frac{Q^*_H}{[\Theta_H]}
+\frac{Q^*_L}{[\Theta_L]}.
\ee
Taking \R{A15}$_3$ into account, we obtain an inequality of the mean values of the contact
temperatures belonging to the cycle
\bee{A17}
0\ \leq\ \Big(-\frac{1}{[\Theta_H]}+\frac{1}{[\Theta_L]}\Big)Q^*_L
\quad\longrightarrow\quad [\Theta_H]\ >\ [\Theta_L]\quad\longrightarrow\quad\eta_{Car}\ \geq\
\eta_{neq}\ >\ 0,
\ee
and the positive definiteness of $\eta_{neq}$ which is according to \R{A17}$_3$ a more realistic
efficiency in comparison with that of Carnot.

\end{document}